\documentclass[final]{siamltex} 
\usepackage{amsmath, amssymb, amsfonts}
\usepackage[export]{adjustbox}
\usepackage{graphicx,color}
\usepackage{diagbox}
\usepackage{algorithm}
\usepackage{algpseudocode}
\usepackage{enumitem}
\usepackage{multirow}
\usepackage{subfig}
\usepackage{epstopdf}
\usepackage{comment}
\usepackage{adjustbox}
\usepackage{booktabs}
\usepackage{siunitx}

\newcommand{\bvec}[1]{\mathbf{#1}}

\renewcommand{\Im}{\mathrm{Im}}

\newcommand{\mc}[1]{\mathcal{#1}}

\newcommand{\vr}{\bvec{r}}

\newcommand{\vR}{\bvec{R}}
\newcommand{\vq}{\bvec{q}}

\newcommand{\ud}{\,\mathrm{d}}

\newcommand{\abs}[1]{\lvert#1\rvert}

\newcommand{\wt}[1]{\widetilde{#1}}

\newcommand{\ie}{\textit{i.e.}~}

\newcommand{\Or}{\mathcal{O}}

\newcommand{\I}{\mathrm{i}}

\newcommand{\RR}{\mathbb{R}}
\newcommand{\CC}{\mathbb{C}}

\newcommand{\vk}{\bvec{k}}
\newcommand*{\conj}[1]{\overline{#1}}
\newcommand{\bpsi}{\bar{\psi}}
\newcommand{\hbse}{H_\text{BSE}}
\newcommand{\eff}{\text{eff}}
\newcommand{\hr}[1]{\hat{\vr}_{#1}}

\numberwithin{equation}{section}
\numberwithin{figure}{section}

\providecommand{\corollaryname}{Corollary}
\providecommand{\lemmaname}{Lemma}
\providecommand{\propositionname}{Proposition}
\providecommand{\remarkname}{Remark}
\providecommand{\theoremname}{Theorem}

\newcommand{\va}{\bvec{a}}
\newcommand{\vb}{\bvec{b}}
\newcommand{\vG}{\bvec{G}}

\title{Fast optical absorption spectra calculations \\for periodic solid state  systems}
\author{Felix Henneke\thanks{Institut f\"ur Mathematik, Freie Universit\"at Berlin, Germany, Email: \texttt{felix.henneke@fu-berlin.de}} \and
Lin Lin\thanks{Department of Mathematics, University of California, Berkeley, and Computational Research Division, Lawrence Berkeley National Laboratory, Berkeley, CA 94720. Email: \texttt{linlin@math.berkeley.edu}} \and
Christian Vorwerk\thanks{Institut f\"ur Physik and IRIS Adlershof, Humboldt-Universit\"at zu Berlin, Germany, Email: \texttt{vorwerk@physik.hu-berlin.de}} \and
Claudia Draxl\thanks{Institut f\"ur Physik and IRIS Adlershof, Humboldt-Universit\"at zu Berlin, Germany, Germany, Email: \texttt{claudia.draxl@physik.hu-berlin.de}}  \and
Rupert Klein\thanks{Institut f\"ur Mathematik, Freie Universit\"at Berlin, Germany, Email: \texttt{rupert.klein@fu-berlin.de}}  \and
Chao Yang\thanks{Computational Research Division, Lawrence Berkeley National Laboratory, Berkeley, CA 94720. Email: \texttt{cyang@lbl.gov}}}

\begin{document}

\maketitle

\begin{abstract}
We present a method to construct an efficient approximation to 
the bare exchange and screened direct interaction kernels of the 
Bethe-Salpeter Hamiltonian for periodic solid state systems via the 
interpolative separable density  fitting technique.
We show that the cost of constructing the approximate Bethe-Salpeter Hamiltonian
scales nearly optimally as $\mathcal{O}(N_k)$ with respect to the number of samples in the Brillouin zone $N_k$. In addition, we show that the cost for applying the Bethe-Salpeter Hamiltonian to a vector scales as $\mathcal{O}(N_k \log N_k)$. Therefore the 
optical absorption spectrum, as well as selected excitation energies  can be efficiently computed via iterative methods such as the Lanczos method. This is a significant reduction from the $\mathcal{O}(N_k^2)$ and $\mathcal{O}(N_k^3)$ scaling
associated with a brute force approach for constructing the Hamiltonian and diagonalizing the Hamiltonian respectively. 
We demonstrate the efficiency and accuracy of this approach with both 
one-dimensional model problems and three-dimensional real materials (graphene and diamond). For the diamond system with $N_k=2197$, it takes $6$ hours to assemble the Bethe-Salpeter Hamiltonian and $4$ hours to fully diagonalize the Hamiltonian using $169$ cores when the brute force approach is used.  The new method takes less than $3$ minutes to set up the Hamiltonian and $24$ minutes to compute the absorption spectrum on a single core.
\end{abstract}

\begin{keywords}
  Bethe--Salpeter equation, interpolative separable density fitting, optical absorption function
\end{keywords}

\begin{AMS}
  65F15, 65Z05
\end{AMS}

\section{Introduction}

The Bethe--Salpeter equation (BSE), derived from the many-body perturbation theory (MBPT), is a widely used method for describing the optical absorption process in molecules and solids~\cite{PRB_62_4927_2000,PR_84_1232_1951_BSE,Strinati1988,Onida1995,Albrecht1997,RMP_74_601_2002,CPC_183_1269_2012_BerkeleyGW}. It models the behavior of an electron--hole pair, which is an excitation process with two quasi-particles.  
Solving BSE requires constructing and diagonalizing a structured
matrix, called the Bethe--Salpeter Hamiltonian (BSH). In the context of optical absorption, the eigenvalues of the BSH are the exciton energies and the corresponding eigenfunctions yield the exciton wavefunctions.
The BSH consists of the so called bare exchange and screened direct interaction kernels
that depend on single-particle orbitals obtained from a
quasi-particle (usually at the GW level) or mean-field calculation.
For isolated systems such as molecules,
the construction of these kernels requires at least $\Or(N_e^5)$
operations in a conventional approach, where $N_e$ is the number of electrons 
in the system. This is very costly for large systems that contain hundreds or more atoms.
Recent efforts have actively explored methods for efficient representation of the BSH, in order to reduce the high computational cost of BSE
calculations~\cite{JCP_334_221_2017,MolPhys_114_1148_2016,PRB_92_075422_2015,PRB_95_075415_2017,RoccaLuGalli2010,PingRoccaGalli2013,PingRoccaLuEtAl2012,RoccaPingGebauerEtAl2012}.

In a recent work~\cite{ICCS}, two of the authors have presented an efficient way to construct the BSH for molecular systems, and to efficiently solve the BSE eigenvalue problem using an iterative scheme. Our approach is based on the recently-developed interpolative separable density fitting (ISDF)
decomposition~\cite{LuYing2015,LuYing2016}. The ISDF decomposition has
been applied to accelerate a number of applications in computational
chemistry and materials science, including the computation of two-electrons
integrals~\cite{LuYing2015}, correlation energy in the
random phase approximation~\cite{LuThicke2017}, density functional
perturbation theory~\cite{LinXuYing2017}, and hybrid density
functional calculations~\cite{HuLinYang2017,DongHuLin2018}.
In this scheme, a matrix consisting of products of single-particle orbital pairs
is efficiently approximated as a low-rank matrix product, between a matrix built with a small number of auxiliary basis vectors
and an expansion coefficient matrix.
This decomposition  allows us to construct efficient representations to the bare exchange and screened direct
kernels. For isolated systems, the construction of the ISDF-compressed BSH matrix only requires 
$\Or(N_{e}^3)$ operations when the rank of the numerical
auxiliary basis is kept at $\Or(N_{e})$. This results in considerate reduction of the cost compared to the $\Or(N_{e}^5)$ complexity required in a
conventional approach.  By keeping the interaction kernels in a decomposed form, the matrix--vector multiplications required in the iterative diagonalization procedures of
the Hamiltonian $\hbse$ can be performed efficiently.
We can further use these efficient matrix--vector
multiplications in a structure preserving Lanczos
algorithm~\cite{ShaoJornadaLinEtAl2018} to obtain an approximate
absorption spectrum without an explicit diagonalization of the
approximate $\hbse$. 

This paper generalizes the work in ~\cite{ICCS} to periodic solid state systems. 
According to the Bloch decomposition, each single particle orbital in a periodic system can be characterized by an orbital index $i$, and a Brillouin zone index $\vk$. Compared to isolated systems, the total number of electrons $N_e$ is equal to the number of electrons per unit cell multiplied by the number of $\vk$ points denoted by $N_k$.  It has been observed that for many extended systems, the number of orbitals (both occupied and virtual orbitals) required for one particular $\vk$ index can be relatively small, and is independent of $N_e$. Hence the difficulty of optical absorption spectra calculations for periodic systems mainly arise from the large number of $\vk$-points. This is particularly the case when the excitons are delocalized in the real space, or when the Fermi-surface is not smooth (such as graphene, and other metallic systems). In such case, $N_k$ can often be rather large (from hundreds to hundreds of thousands, see e.g.~\cite{QiuFelipeLouie2013}, where a $120\times 120\times 1$ $\vk$-grid is used for the quasi two-dimensional MoS$_2$ system) in order to properly discretize and sample the Brillouin zone.  The cost for constructing the bare exchange and screened direct kernels scales as $\Or(N_k^2)$, while the cost for diagonalizing the corresponding BSH scales as $\Or(N_k^3)$. This is prohibitively expensive when a dense discretization 
of the Brillouin zone is needed.

With the help of ISDF, we can find a reduced representation of the pair product orbitals in the periodic setting~\cite{LuYing2016}. Such a reduced representation is possible,  thanks to the smoothness of the single particle orbitals with respect to the $\vk$ index, and that the Brillouin zone is a compact domain. We will show that 
we can reduce the complexity of the bare exchange and screened direct kernel construction for extended systems to the optimal complexity of $\mathcal{O}(N_k)$. Instead of diagonalizing the BSH directly, we use iterative methods such as the Lanczos method to evaluate the optical absorption spectrum. The complexity of applying the approximated kernels to a vector with respect to $N_k$ is only $\mathcal{O}(N_k \log N_k)$. The same strategy can be applied to evaluate selected excitation energies. 

The rest of the paper is organized as follows. We first provide a concise review of the single particle theory and the Bethe-Salpeter equation for periodic systems in section~\ref{sec:prelim}. We could not find a precise mathematical description of how the BSH is constructed for periodic systems with a discretized Brillouin zone in the literature. We therefore provide a self-contained derivation in section~\ref{sec:bse_period}. 
The interpolative separable density fitting for periodic systems is introduced in section~\ref{sec:isdf}, and the application of the approximate BSH in the ISDF format to a vector in section~\ref{sec:BSH_apply}. The numerical results are presented in section~\ref{sec:numer}, followed by a conclusion in section~\ref{sec:conclude}.

\section{Preliminaries}\label{sec:prelim}

\subsection{Single particle theory for periodic systems}

To facilitate further discussion we briefly review
Bloch-Floquet theory for periodic systems. Without loss of generality
we consider a three-dimensional crystal.
The \emph{Bravais lattice} with lattice vectors $\va_{1},\va_{2},\va_{3}\in \RR^{3}$ is defined as
\begin{equation}
  \mathbb{L} = \left\{ \vR \vert \vR = n_1\bvec{a}_1 + n_2 \bvec{a}_2 +
n_3\bvec{a}_3, \quad n_1,n_2,n_3\in\mathbb{Z}\right\}.
  \label{eqn:Blattice}
\end{equation}
In single particle theories such as the Kohn-Sham density functional theory, the self-consistent effective potential
$V_{\eff}$ is real-valued and $\mathbb{L}$-periodic, \ie~
\[
V_{\eff}(\vr+\vR) = V_{\eff}(\vr), \quad \forall \vr\in \RR^3, \vR\in
\mathbb{L}.
\]
The unit cell is defined as
\begin{equation}
 \Omega =\left\{ \vr = c_{1} \va_{1} + c_{2} \va_{2} + c_{3}
 \va_{3} ~\vert~ 0 \le c_{1},c_{2},c_{3} < 1\right\}.
 \label{eqn:unit_cell}
\end{equation}
The Bravais lattice induces a reciprocal lattice
$\mathbb{L}^{*}$, with its lattice vectors $\bvec{b}_1,\bvec{b}_2,\bvec{b}_3$
satisfying $\bvec{a}_{\alpha} \cdot \bvec{b}_{\beta} = 2\pi
\delta_{\alpha\beta}, \alpha,\beta \in \{1,2,3\}$.
The unit cell of the reciprocal lattice is called the (first) Brillouin zone and denoted by
$\Omega^{*}$, defined as
\[
\Omega^{*} = \left\{ \vk = k_{1} \vb_{1} + k_{2} \vb_{2} + k_{3}
  \vb_{3} ~\Big\vert~ -\frac12 \le k_{1},k_{2},k_{3} < \frac12\right\}.
\]
The Brillouin zone has a number of special points related to the
symmetry of the crystal. The common special point is the
$\Gamma$-point, which corresponds to $\vk=(0,0,0)^\top$.

According to the Bloch-Floquet theory, the spectrum of the Hamiltonian $\mc{H} = -\frac{1}{2} \nabla_{\vr}^2 + V_\eff(\vr)$ can be
relabeled using two indices $(i,\vk)$, where $i\in \mathbb{N}$ is called the band
index and $\vk\in\Omega^{*}$ is the Brillouin zone index.
Each generalized eigenfunction $\psi_{i\vk}(\vr)$ is known as a Bloch orbital and
satisfies
$\mc{H}\psi_{i\vk}(\vr)=\epsilon_{i\vk}\psi_{i\vk}(\vr)$ with Bloch
boundary conditions $\psi_{i\vk}(\vr + \vR) = e^{\I \vk \cdot \vR}\psi_{i\vk}(\vr)$ for any $\vR\in\mathbb{L}$. 
Furthermore,
$\psi_{i\vk}$ can be decomposed using the Bloch decomposition
\begin{equation}
  \psi_{i\vk}(\vr) = e^{\I \vk \cdot \vr} u_{i\vk}(\vr),
  \label{eqn:Blochdecompose}
\end{equation}
where $u_{i\vk}(\vr)$ is the periodic part of $\psi_{i\vk}(\vr)$ satisfying the periodic boundary condition on the unit cell
\begin{equation}
  u_{i\vk}(\vr+\vR) = u_{i\vk}(\vr), \quad \forall \vR\in\mathbb{L}.
  \label{eqn:upbc}
\end{equation}
It can be directly obtained by solving the eigenvalue problem
\begin{equation}
  \mc{H}(\vk) u_{i\vk} = \epsilon_{i\vk} u_{i\vk}(\vr), \quad \vr\in \Omega, \quad \vk \in
  \Omega^{*},
  \label{eqn:bandproblem}
\end{equation}
where $\mc{H}(\vk) = -\frac12 (\nabla_{\vr}+\I \vk)^2 + V_{\eff}(\vr)$.
For each $\vk\in\Omega^{*}$, the eigenvalues $\epsilon_{i\vk}$ are ordered
non-decreasingly. For a fixed $i$, $\{\epsilon_{i\vk}\}$ as a
function of $\vk$ is called a
\emph{Bloch band}. The collection of all eigenvalues forms the
\emph{band structure} of the crystal, which characterizes the spectrum
of the operator $\mc{H}$.

In the discussion below, we denote by $N_{v}$ the number of valence
bands (i.e., occupied orbitals per unit cell in the ground state), $N_{c}$ the number of
conduction bands (i.e. unoccupied orbitals per unit cell in the ground state). We also define $N=N_v+N_c$. 
We assume the systems to be insulating, in the sense that the
following band isolation conditions between the valence and conduction bands are satisfied:
\begin{equation}
  \inf \abs{\epsilon_{i\vk}-\epsilon_{i'\vk'}}:=\epsilon_{g}>0, \quad
  \vk,\vk'\in\Omega^*, \ \ 1\le i\le N_{v}, \ \ N_{v}+1\le i' \le
  N.
  \label{eqn:gapcondition}
\end{equation}

Denote by $\abs{\Omega}$ the volume of the unit cell, and
\[
\abs{\Omega^{*}} = \frac{(2\pi)^{3}}{\abs{\Omega}}
\]
the volume of the Brillouin zone.
The Bloch orbitals $\{\psi_{i\vk}\}$ satisfy the orthonormality
condition in the distributional sense
\begin{equation}
  \int_{\RR^3} \psi_{i'\vk'}^{*}(\vr) \psi_{i,\vk}(\vr) \ud \vr
  = \abs{\Omega^{*}} \, \delta_{i',i} \, \delta(\vk'-\vk).
  \label{eqn:psiorthogonal}
\end{equation}
Here  $\delta_{i',i}$ is the Kronecker $\delta$ symbol 
for a discrete set, while $\delta(\vk'-\vk)$ is the Dirac-delta distribution.
Equation \eqref{eqn:psiorthogonal} implies the normalization condition when integrated over the
Brillouin zone
\begin{equation}
  \frac{1}{\abs{\Omega^{*}}}\int_{\Omega^{*}} \int_{\RR^3} \psi_{i'\vk}^{*}(\vr)
  \psi_{i\vk}(\vr) \ud \vr \ud \vk = \delta_{i',i}.
  \label{}
\end{equation}
From the Bloch orbitals, the ground state electron density can be constructed as
\begin{equation}
  \rho(\vr) = \frac{1}{|\Omega^{*}|} \int_{\Omega^{*}}
  \sum_{i=1}^{N_{v}} \abs{\psi_{i\vk}(\vr)}^2 \ud \vk =
  \frac{1}{|\Omega^{*}|} \int_{\Omega^{*}} \sum_{i=1}^{N_{v}} \abs{u_{i\vk}(\vr)}^2 \ud \vk.
  \label{eqn:density}
\end{equation}

In order to practically perform calculations for periodic systems, the
integration with respect to the Brillouin zone $\Omega^{*}$ needs to be
discretized using a quadrature. The most commonly used scheme
is based on the Monkhorst-Pack grid~\cite{MonkhorstPack1976}
\begin{equation}
  \mc{K}^{\ell}_{\bvec{s}} = \left\{\sum_{\alpha=1}^{3} \frac{m_{\alpha}-s_{\alpha}}{N^{\ell}_{\alpha}}
\bvec{b}_{\alpha}\;\Big\vert\;
m_{\alpha}=-\frac{N^{\ell}_{\alpha}}{2}+1,\ldots,\frac{N^{\ell}_{\alpha}}{2},\quad
0\le s_{\alpha}< 1,
\quad \alpha=1,2,3\right\}.
  \label{eqn:mpgrid}
\end{equation}
It is clear that $\mc{K}^{\ell}_{\bvec{s}}\subset \Omega^{*}$ and that it
corresponds to a uniform discretization of the Brillouin zone.
When the shift vector $\bvec{s}=\bvec{0}$, we denote by
$\mc{K}^{\ell}:=\mc{K}^{\ell}_{\bvec{0}}$, and the calculation of
periodic systems can be \textit{equivalently} performed using a
supercell consisting of
$N^{\ell}_{1}\times N^{\ell}_{2} \times N^{\ell}_{3}$ unit
cells. The supercell is denoted by $\Omega^{\ell}$, and is
further equipped with periodic boundary condition called the Born-von Karman boundary
condition~\cite{AshcroftEtAl1976}.
The calculation of a periodic crystal can thus be
recovered by taking the limit $N^{\ell}_{\alpha}\to \infty$. We denote
by $N_{k}\equiv N^{\ell}:=N^{\ell}_{1}N^{\ell}_{2}N^{\ell}_{3}$ the total number of
unit cells, or equivalently the total number of Monkhorst-Pack grid
points in the Brillouin zone.

Assuming the Brillouin zone is discretized using $\mc{K}^{\ell}$, the
orthogonality condition~\eqref{eqn:psiorthogonal} becomes
\begin{equation}
     \int_{\Omega^{\ell}} \psi_{i'\vk'}^{*}(\vr) \psi_{i\vk}(\vr) \ud \vr
     = \delta_{i',i} \, \delta_{\vk',\vk}, \quad \vk,\vk'\in \mc{K}^{\ell}.
  \label{eqn:psiorthogonal_discrete}
\end{equation}
We also modify the Bloch decomposition as
\begin{equation}
  \psi_{i\vk}(\vr) = \frac{1}{\sqrt{N^{\ell}}} e^{\I
  \vk\cdot \vr} u_{i\vk}(\vr), \quad \vk\in \mc{K}^{\ell}.
    \label{eqn:blochdecompose_discrete}
\end{equation}
Here the normalization factor $1/\sqrt{N^{\ell}}$ is introduced so that
the orthogonality condition  for the periodic part implies
\begin{equation}
  \int_{\Omega} u_{i'\vk}^{*}(\vr) u_{i\vk}(\vr)\ud \vr  = \delta_{i',i}, \quad \vk\in \mc{K}^{\ell}.
  \label{eqn:uorthogonal_discrete}
\end{equation}

To facilitate the book-keeping effort of various relevant constants in
practical calculations, in the discussion below we will always assume
that the Brillouin zone is discretized into $\mc{K}^{\ell}$ with a
corresponding supercell $\Omega^{\ell}$. The volume of the supercell is
$\abs{\Omega^{\ell}} = N^{\ell} \abs{\Omega} = N_{k} \abs{\Omega}$. The unit cell is
further discretized into a uniform grid $\{\vr_{i}\}_{i=1}^{N_{g}}$.
Practical BSE calculations often truncate
the number of conduction bands aggressively, in the sense that $N_g\gg
N_{v}+N_{c} =: N$. Numerical results indicate that in many cases, the low-lying excitation spectrum 
is relatively insensitive to $N_c$, and one can often choose $N_c\approx N_v$.
Unless otherwise clarified, we may not distinguish a continuous vector
$u(\vr)$ and the corresponding discretized vector $\{u(\vr_{i})\}$.  Similarly, when the context is clear, 
we do not
distinguish the kernel of an operator $A(\vr,\vr')$ and its discretized
matrix $\{A(\vr_{i},\vr_{j})\}$.

\subsection{Bethe-Salpeter equation for periodic systems}\label{sec:bse_period}

The Bethe--Salpeter equation is an eigenvalue problem of the form
\begin{equation}
H_\text{BSE} X = E X \;, \label{eq:BSE}
\end{equation}
where $H_\text{BSE}$ is the Bethe--Salpeter Hamiltonian (BSH), $X$ is the exciton wavefunction, and $E$ is the corresponding exciton
energy.  For periodic systems, the BSH has
the following block structure
\begin{equation}
   H_\text{BSE} =
  \begin{bmatrix}
   D + 2V_A - W_A &  2V_B - W_B \ \\
   -2\conj{V}_B + \conj{W}_B & - D -2\conj{V}_A + \conj{W}_A \ \\
  \end{bmatrix},
\label{eq:BSEHamiltonian}
\end{equation}
where $D(i_v i_c \vk,j_vj_c \vk') = (\epsilon_{i_c \vk} -
\epsilon_{i_v \vk})\delta_{i_v,j_v} \delta_{i_c,j_c} \delta_{\vk,\vk'}$ is an
$(N_vN_c N_{k})\times(N_vN_c N_{k})$ diagonal matrix.
The quasi-particle energies $\epsilon_{i_{v}\vk},\epsilon_{i_{c}\vk}$ are
typically obtained from a GW calculation~\cite{PRB_62_4927_2000}. The $V_A$ and $V_B$ matrices
represent the bare {\em exchange} interaction of electron--hole pairs, and the
$W_A$ and $W_B$ matrices are referred to as the screened {\em
direct} interaction of electron--hole pairs. These matrices are
defined as follows:
\begin{equation}
\begin{split}
  V_A(i_vi_c\vk,j_vj_c \vk') &=\int_{\Omega^{\ell}\times \Omega^{\ell}} \bpsi_{i_c\vk}(\vr)\psi_{i_v\vk}(\vr)
V(\vr,\vr')\bpsi_{j_v\vk'}(\vr')\psi_{j_c\vk'}(\vr')\,\mathrm{d}\vr\,\mathrm{d}\vr',
\\
V_B(i_vi_c\vk,j_vj_c\vk') &=\int_{\Omega^{\ell}\times \Omega^{\ell}} \bpsi_{i_c\vk}(\vr)\psi_{i_v\vk}(\vr)
V(\vr,\vr')\bpsi_{j_c\vk'}(\vr')\psi_{j_v\vk'}(\vr')\,\mathrm{d}\vr\,\mathrm{d}\vr',
\\
W_A(i_vi_c\vk,j_vj_c\vk')
&=\int_{\Omega^{\ell}\times \Omega^{\ell}} \bpsi_{i_c\vk}(\vr)\psi_{j_c\vk'}(\vr)W(\vr,\vr')
\bpsi_{j_v\vk'}(\vr')\psi_{i_v\vk}(\vr') \,\mathrm{d} \vr
\,\mathrm{d}\vr', \\
 W_B(i_vi_c\vk,j_vj_c\vk')
&=\int_{\Omega^{\ell}\times \Omega^{\ell}} \bpsi_{i_c\vk}(\vr)\psi_{j_v\vk'}(\vr)W(\vr,\vr')
\bpsi_{j_c\vk'}(\vr')\psi_{i_v\vk}(\vr') \,\mathrm{d} \vr
\,\mathrm{d}\vr'.
\end{split}
\label{eq:BSEreal}
\end{equation}
Here $\psi_{i_{v}\vk}$ and $\psi_{i_{c}\vk}$ are the
valence and conduction single-particle orbitals typically obtained from a
Kohn--Sham density functional theory (KSDFT) calculation
respectively, and $V(\bvec{r},\bvec{r'})$ and
$W(\bvec{r},\bvec{r'})$ are the bare and screened Coulomb interactions.
Both $V_A$ and $W_A$ are Hermitian, whereas $V_B$ and $W_B$ are
complex symmetric. Within the so-called Tamm--Dancoff
approximation (TDA)~\cite{RMP_74_601_2002}, both $V_B$ and $W_B$ are
neglected in Equation~\eqref{eq:BSEHamiltonian}. In this case, the $\hbse$
becomes Hermitian and we can focus on computing the upper left block of $H_\text{BSE}$.

In the following discussion, when a single index $i$ is used, it refers
to either  $i_{v}$ or $i_{c}$.
Using the Bloch decomposition~\eqref{eqn:blochdecompose_discrete}, the
matrix elements of the BSH can be written using the periodic
part of the orbitals as
\begin{equation}
  \begin{split}
    V_A(i_vi_c\vk,j_vj_c \vk') &=
    \frac{1}{N_{k}^2}
    \int_{\Omega^{\ell}\times \Omega^{\ell}}
    \bar{u}_{i_c\vk}(\vr)u_{i_v\vk}(\vr)
    V(\vr,\vr') \bar{u}_{j_v\vk'}(\vr') u_{j_c\vk'}(\vr')\,\mathrm{d}\vr\,\mathrm{d}\vr',
    \\
    V_B(i_vi_c\vk,j_vj_c\vk') &=
    \frac{1}{N_{k}^2}  \int_{\Omega^{\ell}\times \Omega^{\ell}} \bar{u}_{i_c\vk}(\vr)
    u_{i_v\vk}(\vr)
    V(\vr,\vr')\bar{u}_{j_c\vk'}(\vr')u_{j_v\vk'}(\vr')\,\mathrm{d}\vr\,\mathrm{d}\vr',
    \\
    W_A(i_vi_c\vk,j_vj_c\vk')
    &=
    \frac{1}{N_{k}^2}  \int_{\Omega^{\ell}\times \Omega^{\ell}} e^{-\I (\vk-\vk') \cdot (\vr-\vr')}
    \bar{u}_{i_c\vk}(\vr)u_{j_c\vk'}(\vr)W(\vr,\vr')
    \bar{u}_{j_v\vk'}(\vr')u_{i_v\vk}(\vr') \,\mathrm{d} \vr
    \,\mathrm{d}\vr', \\
    W_B(i_vi_c\vk,j_vj_c\vk')
    &=
    \frac{1}{N_{k}^2}
    \int_{\Omega^{\ell}\times \Omega^{\ell}} e^{-\I (\vk-\vk') \cdot (\vr-\vr')}
    \bar{u}_{i_c\vk}(\vr)u_{j_v\vk'}(\vr)W(\vr,\vr')
    \bar{u}_{j_c\vk'}(\vr')u_{i_v\vk}(\vr') \,\mathrm{d} \vr
    \,\mathrm{d}\vr'.
  \end{split}
  \label{eqn:BSEu}
\end{equation}
Note that $V_{A},V_{B}$ in Eq.~\eqref{eqn:BSEu} do not involve the
phase factors, since the factor $e^{\I \vk \cdot \vr}$
exactly cancels due to the complex conjugate operation. The phase
factor only appears in the $W_{A},W_{B}$ terms.

Eq.~\eqref{eqn:BSEu} requires the evaluation of integrals of the
following form
\begin{equation}
  \mc{V}(f,g) := \frac{1}{N_{k}}
  \int_{\Omega^{\ell}\times \Omega^{\ell}}
  \bar f(\vr) V(\vr,\vr') g(\vr')\,\mathrm{d}\vr\,\mathrm{d}\vr',
  \label{eqn:Vintegral}
\end{equation}
and
\begin{equation}
  \mc{W}_{\vq}(f,g) := \frac{1}{N_{k}}
  \int_{\Omega^{\ell}\times \Omega^{\ell}}
  e^{-\I \vq \cdot (\vr-\vr')} \bar f(\vr) W(\vr,\vr') g(\vr')\,\mathrm{d}\vr\,\mathrm{d}\vr'.
  \label{eqn:Wintegral}
\end{equation}
Using such notation,
\begin{equation}
  \begin{split}
  V_A(i_vi_c\vk,j_vj_c \vk') = &\frac{1}{N_{k}} \mc{V}(\bar{u}_{i_{v}\vk}
  u_{i_{c}\vk}, \bar{u}_{j_{v} \vk'} u_{j_{c}\vk'}),\\
  V_B(i_vi_c\vk,j_vj_c \vk') = &\frac{1}{N_{k}} \mc{V}(\bar{u}_{i_{v}\vk}
  u_{i_{c}\vk}, \bar{u}_{j_{c} \vk'} u_{j_{v}\vk'}),\\
  W_A(i_vi_c\vk,j_vj_c \vk') = &\frac{1}{N_{k}} \mc{W}_{\vk-\vk'}(\bar{u}_{j_{c}\vk'}
  u_{i_{c}\vk}, \bar{u}_{j_{v} \vk'} u_{i_{v}\vk}),\\
  W_B(i_vi_c\vk,j_vj_c \vk') = &\frac{1}{N_{k}} \mc{W}_{\vk-\vk'}(\bar{u}_{j_{v}\vk'}
  u_{i_{c}\vk}, \bar{u}_{j_{c} \vk'} u_{i_{v}\vk}).
  \end{split}
  \label{}
\end{equation}

In Eq.~\eqref{eqn:Vintegral},~\eqref{eqn:Wintegral}, $f,g$
are periodic functions in the unit cell, and can be represented using
their Fourier representations. For instance,
\begin{equation}
  f(\vr) = \sum_{\vG\in \mathbb{L}^{*}} \hat{f}(\vG) e^{\I \vG\cdot
  \vr},
  \label{eqn:Fourier_vec}
\end{equation}
and its Fourier coefficients can be computed as
\begin{equation}
  \hat{f}(\vG) = \frac{1}{\abs{\Omega}} \int_{\Omega}  e^{-\I \vG\cdot
  \vr} f(\vr) \ud \vr.
  \label{eqn:Fourier_coef_vec}
\end{equation}
Hence Parseval's identity reads
\begin{equation}
  \int_{\Omega} \bar{f}(\vr) g(\vr) \ud \vr = \abs{\Omega}
  \sum_{\vG\in \mathbb{L}^{*}} \bar{\hat{f}}(\vG) \hat{g}(\vG).
  \label{}
\end{equation}

Both of the kernels $V,W$ satisfy the translation symmetry
\begin{equation}
  V(\vr + \vR,\vr' + \vR) = V(\vr,\vr'), \quad
  W(\vr + \vR,\vr' + \vR) = W(\vr,\vr'), \quad \forall \vR \in
  \mathbb{L}.
  \label{eqn:VWtranslation}
\end{equation}
Eq.~\eqref{eqn:VWtranslation} also defines the values of $V,W$ for
$\vr,\vr'$ beyond the supercell $\Omega^{\ell}$.
The Fourier representation of $V$ takes the form
\begin{equation}
  V(\vr,\vr') = \frac{1}{\abs{\Omega^{\ell}}}
  \sum_{\vk\in \mc{K}^{\ell}} \sum_{\vG,\vG'}
  e^{\I (\vk+\vG)\cdot \vr} \hat{V}_{\vk}(\vG,\vG') e^{-\I (\vk+\vG')\cdot \vr'},
  \label{eqn:Fourier_mat}
\end{equation}
and the Fourier coefficients can be computed as
\begin{equation}
  \hat{V}_{\vk}(\vG,\vG') = \frac{1}{\abs{\Omega^{\ell}}}
  \int_{\Omega^{\ell}\times \Omega^{\ell}} \ud \vr \ud \vr'
  e^{-\I (\vk+\vG)\cdot \vr} V(\vr,\vr') e^{\I (\vk+\vG')\cdot \vr'}
  \label{eqn:Fourier_coef_mat}
\end{equation}
Similarly, the Fourier representation for $W$ can be defined.

It should be noted that the Coulomb kernel $V$ only depends on the distance
between $\vr$ and $\vr'$, i.e. it has further translational symmetry
property that
\begin{equation}
  V(\vr+\vr'',\vr'+\vr'') = V(\vr,\vr'), \quad \forall \vr'' \in
  \Omega^{\ell}.
  \label{}
\end{equation}
As a result, its Fourier transform $\hat{V}_{\vk}(\vG,\vG')$ can be
simplified into a diagonal matrix
\begin{equation}
  \hat{V}_{\vk}(\vG,\vG') = \frac{4\pi}{\abs{\vk+\vG}^2} \delta_{\vG,\vG'}.
  \label{eqn:V_fourier}
\end{equation}
In fact, the Coulomb kernel periodized with respect to the supercell
$\Omega^{\ell}$ is defined to be the inverse Fourier transform of
Eq.~\eqref{eqn:V_fourier}.

Using such notation, we have
\begin{equation}
  \begin{split}
    &\int_{\Omega^{\ell}} V(\vr,\vr') g(\vr') \ud \vr'  \\
   =&\frac{1}{\abs{\Omega^{\ell}}} \int_{\Omega^{\ell}} \ud \vr'
   \sum_{\vk\in \mc{K}^{\ell}} \sum_{\vG,\vG'}
   e^{\I (\vk+\vG)\cdot \vr} \hat{V}_{\vk}(\vG,\vG') e^{-\I
   (\vk+\vG')\cdot \vr'} g(\vr')\\
   =&\frac{1}{\abs{\Omega^{\ell}}} \sum_{\vR\in\mathbb{L}}\int_{\Omega} \ud \vr'
   \sum_{\vk\in \mc{K}^{\ell}} \sum_{\vG,\vG'}
   e^{\I (\vk+\vG)\cdot \vr} \hat{V}_{\vk}(\vG,\vG') e^{-\I
   (\vk+\vG')\cdot (\vr'+\vR)} g(\vr'+\vR)\\
   =&\frac{1}{\abs{\Omega^{\ell}}} \int_{\Omega} \ud \vr'
   \sum_{\vk\in \mc{K}^{\ell}} \sum_{\vR\in\mathbb{L}} e^{-\I \vk\cdot\vR} \sum_{\vG,\vG'}
   e^{\I (\vk+\vG)\cdot \vr} \hat{V}_{\vk}(\vG,\vG')
   e^{-\I (\vk+\vG')\cdot \vr'} g(\vr')
 \end{split}
  \label{}
\end{equation}
Here we have used $e^{-\I \vG' \cdot \vR} = 1$, the fact that $g$ is
periodic with respect to the unit cell $\Omega$, as well as the identity
\begin{equation}
  \int_{\Omega^{\ell}} f(\vr') \ud \vr' =
  \sum_{\vR\in\mathbb{L}} \int_{\Omega} f(\vr'+\vR) \ud
  \vr'.
  \label{}
\end{equation}
Furthermore, from Eq.~\eqref{eqn:Fourier_coef_vec} and the identity
\[
\sum_{\vR\in\mathbb{L}} e^{-\I \vk\cdot\vR} = N_k \delta_{\vk, 0}
\]
we have
\begin{equation}
  \begin{split}
   &\int_{\Omega^{\ell}} V(\vr,\vr') g(\vr') \ud \vr'  \\
  =&\frac{1}{\abs{\Omega}} \int_{\Omega} \ud \vr'
   \sum_{\vG,\vG'}
   e^{\I \vG\cdot \vr} \hat{V}_{\bvec{0}}(\vG,\vG')
   e^{-\I \vG'\cdot \vr'} g(\vr')\\
  =& \sum_{\vG,\vG'}
  e^{\I \vG\cdot \vr} \hat{V}_{\bvec{0}}(\vG,\vG') \hat{g}(\vG').
  \end{split}
  \label{}
\end{equation}
Compared to Eq.~\eqref{eqn:V_fourier}, the definition of
$\hat{V}_{\bvec{0}}$ should be modified to
\begin{equation}
  \hat{V}_{\bvec{0}}(\vG,\vG') = \begin{cases}
    \frac{4\pi}{\abs{\vG}^2} \delta_{\vG,\vG'}, \quad & \vG \neq
    \bvec{0},\\
    0,& \vG=\bvec{0}.
  \end{cases}
  \label{eqn:V0Fourier}
\end{equation}
Another way to understand Eq.~\eqref{eqn:V0Fourier} is that it can only
be applied to a mean zero function $g(\vr)$, such that
$\hat{g}(\bvec{0})=0$. In other words, $g$ should be in the range of the
Laplacian operator with the periodic boundary condition.  This is indeed correct for BSE
calculations, due to the orthogonality condition between the valence and
conduction bands
\[
\int_{\Omega} \bar{u}_{i_{c} \vk}(\vr) u_{i_{v}\vk}(\vr) \ud \vr = 0.
\]

This implies
\begin{equation}
  \begin{split}
  \mc{V}(f,g) = &\frac{1}{N_{k}}\int_{\Omega^{\ell}} \bar{f}(\vr) \sum_{\vG,\vG'}
  e^{\I \vG\cdot \vr} \hat{V}_{\bvec{0}}(\vG,\vG') \hat{g}(\vG')\\
  = &\int_{\Omega} \bar{f}(\vr) \sum_{\vG,\vG'}
  e^{\I \vG\cdot \vr} \hat{V}_{\bvec{0}}(\vG,\vG') \hat{g}(\vG')\\
  = &\abs{\Omega} \sum_{\vG,\vG'}
  \bar{\hat{f}}(\vG) \hat{V}_{\bvec{0}}(\vG,\vG') \hat{g}(\vG')\\
  =& \abs{\Omega} \sum_{\vG\ne \bvec{0}} \frac{4\pi}{\abs{\vG}^2}
  \bar{\hat{f}}(\vG) \hat{g}(\vG).
  \end{split}
  \label{eqn:Vintegral_compute}
\end{equation}

Similarly for the $W$ part,
\begin{equation}
  \begin{split}
    &\int_{\Omega^{\ell}} e^{-\I \vq \cdot (\vr-\vr')} W(\vr,\vr') g(\vr') \ud \vr'  \\
   =&\frac{1}{\abs{\Omega^{\ell}}} \int_{\Omega^{\ell}} \ud \vr' e^{-\I \vq \cdot (\vr-\vr')}
   \sum_{\vk\in \mc{K}^{\ell}} \sum_{\vG,\vG'}
   e^{\I (\vk+\vG)\cdot \vr} \hat{W}_{\vk}(\vG,\vG') e^{-\I (\vk+\vG')\cdot \vr'} g(\vr')\\
   =& \frac{1}{\abs{\Omega^{\ell}}} \int_{\Omega} \ud \vr'
   e^{\I (\vk-\vq) \cdot (\vr-\vr')}
   \sum_{\vk\in \mc{K}^{\ell}} \sum_{\vR\in\mathbb{L}} e^{-\I (\vk-\vq)\cdot\vR} \sum_{\vG,\vG'}
   e^{\I \vG\cdot \vr} \hat{W}_{\vk}(\vG,\vG')
   e^{-\I \vG'\cdot \vr'} g(\vr').
  \end{split}
\end{equation}
In order to obtain a non-vanishing quantity in the equation above,
note that the quantity $\sum_{\vR\in\mathbb{L}} e^{-\I (\vk - \vq)\cdot\vR} = N_k$ if $\vk - \vq
\in \mathbb{L}^{*}$, and is otherwise $0$. Therefore the summation with
respect to $\vk$ should be restricted to those satisfying
\[
  \vk-\vq=\vG'',\quad \vG''\in \mathbb{L}^{*}.
\]
Since $\vk$ is restricted to the
first Brillouin zone, there is a unique $\vG''$ (and therefore $\vk$) for each
given $\vq$ satisfying this relation. Also note that $\vk-\vq$ may exceed the first Brillouin
zone. In other words, it is indeed possible to have $\vG''\ne \bvec{0}$.  Then
for a given $\vq$,
\begin{equation}
  \begin{split}
    &\int_{\Omega^{\ell}} e^{-\I \vq \cdot (\vr-\vr')} W(\vr,\vr') g(\vr') \ud \vr'  \\
   =& \frac{1}{\abs{\Omega}} \int_{\Omega} \ud \vr'
   \sum_{\vG,\vG'} e^{\I (\vG+\vG'')\cdot \vr} \hat{W}_{\vG''+\vq}(\vG,\vG')
   e^{-\I (\vG'+\vG'')\cdot \vr'} g(\vr')\\
   =& \sum_{\vG,\vG'} e^{\I (\vG+\vG'')\cdot \vr} \hat{W}_{\vG''+\vq}(\vG,\vG')
   \hat{g}(\vG'+\vG'') \\
   =& \sum_{\vG,\vG'} e^{\I \vG\cdot \vr}
   \hat{W}_{\vG''+\vq}(\vG-\vG'',\vG'-\vG'')
   \hat{g}(\vG')\\
   =& \sum_{\vG,\vG'} e^{\I \vG\cdot \vr}
   \hat{W}_{\vq}(\vG,\vG') \hat{g}(\vG').
  \end{split}\label{eqn:Wfourier}
\end{equation}
In the last equality, we have used the definition of the Fourier
coefficients in Eq.~\eqref{eqn:Fourier_coef_mat}. We then readily have
\begin{equation}
 \mc W_{\vq}(f,g) = \abs{\Omega} \sum_{\vG,\vG'} \bar{\hat{f}}(\vG)
 \hat{W}_{\vq}(\vG,\vG') \hat{g}(\vG').
  \label{eqn:Wintegral_compute}
\end{equation}
Therefore, despite that $\mc W_{\vq}(f,g)$ is significantly more complex to
define, the resulting formula in the Fourier representation is
remarkably similar to the form of $\mc{V}(f,g)$.

\section{Interpolative separable density fitting for periodic systems}\label{sec:isdf}

In order to reduce the computational complexity, we seek to minimize the
number of integrals in Equation~\eqref{eq:BSEreal}. We will use the
interpolative separable density fitting decomposition
(ISDF)~\cite{LuYing2015,LuYing2016}. For periodic systems, we first
consider the following general form of decomposition
\begin{equation}
  Z_{i\vk,j\vk'}(\vr):=u_{i\vk}(\vr) \bar{u}_{j\vk'}(\vr) \approx \sum_{\mu=1}^{N_{\mu}}
  \zeta_{\mu}(\vr)
  u_{i\vk}(\hr{\mu}) \bar{u}_{j\vk'}(\hr{\mu}).
  \label{eqn:ISDFgeneral}
\end{equation}
When the unit cell is discretized into a
uniform grid $\{\vr_{n}\}_{n=1}^{N_{g}}$, $Z$ can be viewed as a matrix
with its row index being $\vr$, and the column index being a multi-index
$(i\vk,j\vk')$.  The matrix size is thus $N_{g}\times N^2 N_{k}^2$ (recall that $N=N_{v}+N_{c}$).
For a given $\vr$, $u_{i\vk}(\vr)\bar{u}_{j\vk'}(\vr)$ can be viewed as a
row vector of size $N^2 N_{k}^2$.
The ISDF decomposition then states that all such matrix rows can be
approximately expanded using a linear combination of matrix rows
with respect to a selected set of \textit{interpolation points}
$\{\hr{\mu}\}_{\mu=1}^{N_{\mu}}\subset \{\vr_{i}\}_{i=1}^{N_{g}}$.  
The coefficients of such a linear
combination, or \textit{interpolating vectors}, are denoted by
$\{\zeta_{\mu}(\vr)\}_{\mu=1}^{N_{\mu}}$. Here $N_{\mu}$ can be
interpreted as the numerical rank of the ISDF decomposition. 

The compression of the pair products $u_{i\vk}(\vr)\bar{u}_{j\vk'}(\vr)$
can be understood from the following two limits. First, if only the $\Gamma$ point is used
to sample the Brillouin zone, we find that there are $N_v N_c\sim N^2$ pairs of functions. However,
the number of grid points $N_{g}$ only scales linearly with respect to $N$. 
Hence the numerical rank of the pair products must scale asymptotically as
$\Or(N)$. In fact, when all orbitals are smooth functions, we can expect that the numerical rank $N_{\mu}$ to be much 
lower than $N_g$. This statement has been confirmed by recent analysis~\cite{LuSoggeSteinerberger}. Second, if a large number of $\vk$-points
are used to discretize the Brillouin zone, $N_v,N_c$ are often relatively small, and the number of 
grid points in the unit cell $N_g$ does not increase with respect to $N_k$. Hence as $N_k$ increases, we may also expect that
the numerical rank $N_{\mu}$ will be determined by smoothness of $u$ with respect to $\vr,\vk$, and is asymptotically independent of $N_k$.  This is indeed what we observe in numerical results. Throughout the discussion below, we will focus on the second scenario, i.e. we will explicitly write down the scaling with respect to $N_g,N$ and $N_k$, but we will 
primarily focus on the scaling with respect to $N_k$.

Assume the interpolation points  $\{\hr{\mu}\}_{\mu=1}^{N_{\mu}}$ are
already chosen, the interpolation vectors can be efficiently
evaluated using a least squares method as follows~\cite{HuLinYang2017}.
Using a linear algebra notation, Eq.~\eqref{eqn:ISDFgeneral} can be written as
\begin{equation}
  Z \approx \Theta C,
  \label{eqn:isdflineq}
\end{equation}
Here $\Theta = [\zeta_1, \zeta_2, ..., \zeta_{N_{\mu}}]$ contains the interpolating vectors.
Each column of $C$ indexed by $(i\vk,j\vk')$ is given by
\[ [ u_{i\vk}(\hr{1})\bar{u}_{j\vk'}(\hr{1}), \cdots,
u_{i\vk}(\hr{\mu})\bar{u}_{j\vk'}(\hr{\mu}), \cdots,
u_{i\vk}(\hr{N_{\mu}})\bar{u}_{j\vk'}(\hr{N_{\mu}})]^{\top}.
\]
Eq.~\eqref{eqn:isdflineq} is an over-determined linear system with
respect to the interpolation vectors $\Theta$. The least squares
approximation to the solution is given by
\begin{equation}
\Theta = ZC^* (CC^*)^{-1}. \label{eq:Theta}
\end{equation}

Due to the tensor product structure  of $Z$ and $C$,
the matrix-matrix multiplications $ZC^*$ and
$CC^*$  can be carried out efficiently~\cite{HuLinYang2017}, with computational cost
being $\Or(N_{g} N_{\mu} N N_{k})$ and $\Or(N^2_{\mu} N N_{k})$,
respectively. The cost of inverting the matrix $CC^{*}$ is
$\Or(N_{\mu}^{3})$, and the
overall cost evaluating $\Theta$ is thus bounded by $\Or(N_{g} N_{\mu} N N_{k} +
N_{\mu}^{3} + N_{g} N_{\mu}^{2})$. Hence the cost
scales cubically with respect to the number of electrons in the unit
cell, and linearly with respect to the number of $\vk$ points.

Eq.~\eqref{eqn:ISDFgeneral} is the general form of ISDF. In the BSE
calculations, we may further distinguish whether
$i,j$ should take valence or conduction band indices only, as well as whether
$\vk,\vk'$ can be set to be the same.  For instance, Eq.~\eqref{eqn:BSEu} suggests
that in order to compress $V_{A},V_{B}$, we only need the following ISDF
decomposition:
\begin{equation}
  Z^{V}_{i_{c}i_{v}\vk}(\vr) := u_{i_{c}\vk}(\vr) \bar{u}_{i_{v}\vk}(\vr) \approx
  \sum_{\mu=1}^{N^{V}_{\mu}}
  \zeta^{V}_{\mu}(\vr)
  u_{i_{c}\vk}(\hr{\mu}) \bar{u}_{i_{v}\vk}(\hr{\mu}).
  \label{eqn:ISDF_V}
\end{equation}
Note that the number of columns of the matrix $Z^{V}$ is only
$N_{v}N_{c}N_{k}$, and the
number of fitting functions $N^{V}_{\mu}$ can be chosen to be less than $N_{\mu}$. 
The computation of $W_A,W_B$ requires the general ISDF
format~\eqref{eqn:ISDFgeneral}. 

The interpolations points $\{\hr{\mu}\}_{\mu=1}^{N_{\mu}}$ can be chosen via a QR factorization
with column pivoting (QRCP) method~\cite{GolubVan2013}, with
randomization to reduce the computational cost. We refer readers
to~\cite{LuYing2015,LuYing2016} for details of the randomized QRCP
method for evaluating the interpolation points.
Other methods can also be used as well to find the interpolation points as well, such as the method based on the centroidal Voronoi decomposition (CVT)~\cite{DongHuLin2018}. 

\section{Fast algorithm for applying the BSH to a vector}\label{sec:BSH_apply}

Once the ISDF decomposition is obtained, we may compute the following
matrix elements
\begin{equation}
  \wt{V}_{A,\mu\nu} = \mc{V}(\zeta^V_\mu, \zeta^V_\nu), \quad
  \wt{V}_{B,\mu\nu} = \mc{V}(\zeta^V_\mu, \bar \zeta^V_\nu), \quad
  \mu,\nu=1,\ldots,N_{\mu}^{V},
  \label{eqn:Vtilde}
\end{equation}
and similarly
\begin{equation}
  \wt{W}_{\vq,\mu\nu} = \mc{W}_{\vq}(\zeta_\mu, \zeta_\nu), \quad
  \mu,\nu=1,\ldots,N_{\mu}.
  \label{eqn:Wtilde}
\end{equation}
The expressions in Eq.~\eqref{eqn:BSEu} can then be approximated in the
ISDF format as
\begin{equation}\label{eqn:VWelements}
  \begin{split}
    V_A(i_vi_c\vk,j_vj_c \vk') &\approx \frac{1}{N_{k}} \sum_{\mu, \nu = 1}^{N_\mu^V}
    \bar u_{i_c\vk}(\hat \vr_\mu)u_{i_v\vk}(\hat\vr_\mu)
    \wt{V}_{A,\mu\nu}
    \bar u_{j_v\vk'}(\hat\vr_\nu) u_{j_c\vk'}(\hat\vr_\nu),\\
    V_B(i_vi_c\vk,j_vj_c \vk') &\approx \frac{1}{N_{k}} \sum_{\mu, \nu = 1}^{N_\mu^V}
    \bar u_{i_c\vk}(\hat \vr_\mu)u_{i_v\vk}(\hat\vr_\mu)
    \wt{V}_{B,\mu\nu}
    \bar u_{j_c\vk'}(\hat\vr_\nu) u_{j_v\vk'}(\hat\vr_\nu),\\
    W_A(i_v i_c \vk,j_v j_c \vk') &= \frac{1}{N_{k}} \sum_{\mu, \nu = 1}^{N_\mu}
    \bar u_{i_c\vk}(\hat\vr_\mu)u_{j_c \vk'}(\hat\vr_\mu)
    \wt{W}_{\vk - \vk',\mu\nu}
    \bar u_{j_v\vk'}(\hat\vr_\nu)u_{i_v\vk}(\hat\vr_\nu),\\
    W_B(i_v i_c \vk,j_v j_c \vk') &= \frac{1}{N_{k}} \sum_{\mu, \nu = 1}^{N_\mu}
    \bar u_{i_c\vk}(\hat\vr_\mu)u_{j_v \vk'}(\hat\vr_\mu)
    \wt{W}_{\vk - \vk',\mu\nu}
    \bar u_{j_c\vk'}(\hat\vr_\nu)u_{i_v\vk}(\hat\vr_\nu).
  \end{split}
\end{equation}

In order to use the Fourier representation~\eqref{eqn:Vintegral_compute}
and~\eqref{eqn:Wintegral_compute}, we first need to perform Fourier
transform for $\{\zeta_{\mu}^{V}\}$ and $\{\zeta_{\mu}\}$. Using the
fast Fourier transform (FFT), and assuming that the number of Fourier
coefficients $\vG$ is also $N_{g}$, the computational cost for the
Fourier transform scales as $\Or(N_{\mu}^{V} N_{g}\log N_{g})$ and
$\Or(N_{\mu} N_{g}\log N_{g})$, respectively. The Fourier coefficients
$\hat{V}_{\vk}$ can be obtained analytically, and we assume the
coefficients $\hat{W}_{\vk}$ are already provided from e.g.~a GW
calculation.  The cost for computing $\wt{V}_{A}, \wt{V}_{B}$ using
Eq.~\eqref{eqn:Vintegral_compute} is then $\Or((N_\mu^V)^2 N_g)$.
Similarly the cost for computing all $\wt{W}_{\vq}$ matrices is
$\Or(N_{\mu}^2 N_{g} N_{k})$. In particular, the total cost for the
initial setup stage scales as $\Or(N_k)$ with respect to the number of
$\vk$-points.  

After 
this initial setup stage, each entry of the BSH can be
computed with $\Or((N_\mu^V)^2 + N_\mu^2)$ operations. If the entire BSH
matrix is to be constructed, the cost will be $\Or(N_{\mu}^2 N_{k}^2 N_v^2 N_c^2)$.

Below we demonstrate that if we only aim at applying the Hamiltonian
$H_\text{BSE}$ to an arbitrary vector without ever assembling the full
Hamiltonian, the computational cost can be greatly reduced.

For simplicity, let us focus on the case when the Tamm--Dancoff
approximation (TDA) is used.
Applying the Hamiltonian $H_\text{BSE} = D + 2V_{A} - W_{B}$ to a vector
$X \in \CC^{N_{v}N_{c}N_{k}}$ amounts to evaluating the three terms
\begin{equation}\label{eqn:HtimesX}
  \begin{split}
    [D X](i_v i_c \vk) &= (\epsilon_{i_c \vk} - \epsilon_{i_v \vk'}) X(i_v i_c \vk),\\
    [V_A X](i_v i_c \vk) &= \sum_{j_v, j_c, \vk'} V_A(i_v i_c \vk, j_v j_c \vk') X(j_v j_c \vk'),\\
    [W_A X](i_v i_c \vk) &= \sum_{j_v, j_c, \vk'} W_A(i_v i_c \vk, j_v j_c \vk') X(j_v j_c \vk').
  \end{split}
\end{equation}
Computing the first term for all $(i_v i_c \vk)$ clearly costs $\mc
O(N_v N_c N_k)$ operations. We now show that the second and third term
can also be computed efficiently.

Using~\eqref{eqn:VWelements}, the second term in~\eqref{eqn:HtimesX} can be regrouped as
\begin{multline}
  \frac{1}{N_k}\sum_\mu
  \bar u_{i_c\vk}(\hat \vr_\mu)u_{i_v\vk}(\hat\vr_\mu)
  \Bigg\{\sum_\nu \wt{V}_{A,\mu\nu} \\
  \left(\sum_{\vk'}\left(\sum_{j_c} u_{j_c\vk'}(\hat\vr_\nu) \left(\sum_{j_v} \bar
  u_{j_v\vk'}(\hat\vr_\nu) X(j_v j_c
  \vk')\right)\right)\right)\Bigg\}.
\end{multline}
This means that one can first perform contractions over $j_v$, $j_c$, and $\vk'$ to obtain a quantity which only depends on $\hat \vr_\nu$.
The computational complexity is $\Or(N_\mu^V (N_v N_c N_k + N_c N_k))$.
The two remaining sums can be computed with $\Or((N_\mu^V)^2 + N_\mu^V
N_v N_c N_k)$ operations.
The total complexity of computing $V_A X$ is bounded by $\Or((N_\mu^V)^2 + N_\mu^V N_v N_c N_k)$.

For the third term in~\eqref{eqn:HtimesX} we obtain
\begin{multline}
  \frac{1}{N_k}\sum_\nu u_{i_v\vk}(\hat\vr_\nu)
  \Bigg\{\sum_\mu \bar u_{i_c\vk}(\hat\vr_\mu) \\
  \left(\sum_{\vk'} \wt{W}_{\vk - \vk',\mu\nu}
  \left(\sum_{j_c} u_{j_c\vk'}(\hat\vr_\mu) \left(\sum_{j_v} \bar u_{j_v
  \vk'}(\hat\vr_\nu) X(j_v j_c \vk')\right)\right)\right)\Bigg\}.
\end{multline}
Here, the two innermost contractions over $j_v$ and $j_c$ result in a
quantity that only depends on $\vk$, $\hat \vr_\mu$, and $\hat \vr_\nu$.
The cost for these two steps is $\Or(N_\mu N_k N_v N_c + N_\mu^2 N_k N_c)$.
The sum over $\vk'$ has the structure of a \textit{discrete
convolution}, for each fixed $\mu\nu$ pair. Therefore it can be computed
for all $\vk$ simultaneously in $\Or(N_\mu^2 N_k \log N_k)$ operations
by fast convolution algorithms, e.g., by using FFT with zero-padded vectors.
The remaining summation operations over $\mu$ and $\nu$ are then obtained with $\Or(N_\mu^2 N_c N_k + N_\mu N_v N_c N_k)$ operations.
In total the computation of $W_A X$ amounts to $\Or(N_\mu N_v N_c N_k
+ N_\mu^2 N_c N_k + N_\mu^2 N_k \log N_k)$ operations.

Combining the results for the three parts of the Hamiltonian, we see
that the computational complexity is given by
\[
\Or((N_\mu + N_\mu^V) N_v
N_c N_k + (N_\mu^V)^2  + N_\mu^2 N_c N_k + N_\mu^2 N_k \log N_k).
\]
In particular, the cost with respect to the number of $\vk$ points only
scales as $\Or(N_k \log N_k)$. This allows us to perform BSE calculations for
complex materials which requires a very large number of $\vk$-points.

By avoiding the explicit construction of $H_\text{BSE}$,
the new algorithm also drastically reduces the storage cost. The
storage cost for $H_\text{BSE}$ alone is $\Or((N_{v}N_{c}N_{k})^2)$.
In the new algorithm, the storage cost of
$\hat{W}_{\vq}$ becomes the dominant component and scales only linearly
with respect to $N_{k}$.

As an example, the matrix-free application of $H_\text{BSE}$ can be used
to compute the optical absorption spectrum, which requires
the evaluation of the following quantity
\begin{equation}
  \varepsilon_2(\omega) = \Im\biggl[ \frac{8\pi}{\abs{\Omega}}
d_r^*\bigl((\omega-\I\eta){I}-H_\text{BSE}\bigr)^{-1}d_l
\biggr], \label{eqn:absorption}
\end{equation}
Here $d_r$ and $d_l$ are called the right and left optical
transition vectors, and $\eta$ is a broadening factor used to
account for the exciton lifetime.
We also compute the smallest eigenvalue of $H_\text{BSE}$ which are of interest in their own right, as they represent the transition energies of bound excitons in many semiconducting solid state materials.

To observe the absorption spectrum and identify its main peaks,
it is possible to use a structure preserving iterative
method instead of explicitly computing all eigenpairs of
$H_\text{BSE}$. We refer readers to
Ref.~\cite{JCTC_11_5197_2015,ShaoJornadaLinEtAl2018} for details of the
structure preserving Lanczos algorithm, which has been implemented in the
BSEPACK~\cite{BSEPACK_UserGuide_2016} library.
When TDA is used, the structure preserving Lanczos reduces to a standard
Lanczos algorithm. 
For the computation of the first eigenvalue we use standard ARPACK~\cite{arpack} routines for Hermitian matrices. 

\section{Numerical Examples}\label{sec:numer}

To illustrate the efficiency of ISDF for BSE calculations in crystals, we apply the method to compute the excitation modes and absorption spectra of 
a one-dimensional model problem as well as two real material systems, diamond (3D bulk) and graphene (quasi-2D).
For both systems, we determine the optical absorption spectra on $\vk$-grids close to those employed in previously published calculations to 
demonstrate that our method is suitable for state-of-the-art calculations, both for 3D and quasi-2D materials.
We furthermore provide a numerical scaling analysis and a more detailed analysis of the error in the ISDF in the case of the one-dimensional 
model and diamond. We show that a good approximation of the spectrum can be obtained with a small number of interpolation vectors.

The method was implemented in \texttt{Julia}~\cite{julia} and the source code is available at \texttt{github.com/fhenneke/BSE\_k\_ISDF.jl}.
As input to our method for the actual materials, we employ 
the KSDFT single-particle orbitals, quasi-particle energies
and screened Coulomb potential computed by \texttt{exciting}~\cite{exciting, Vorwerk2019}, an all-electron full-potential code with
implementations of density functional theory and many-body perturbation theory.
The Tamm--Dancoff approximation is used in all calculations.

All calculation for the proposed method were carried out on a single core of an i5-8250U CPU at 1.60GHz.

\subsection{One-dimensional problems}

\begin{figure}
\includegraphics[valign=t]{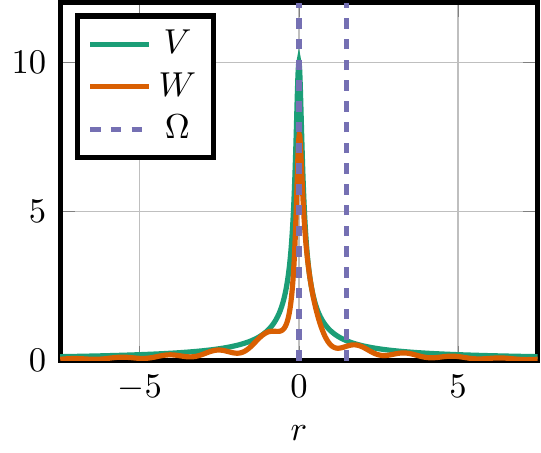}
\quad
\includegraphics[valign=t]{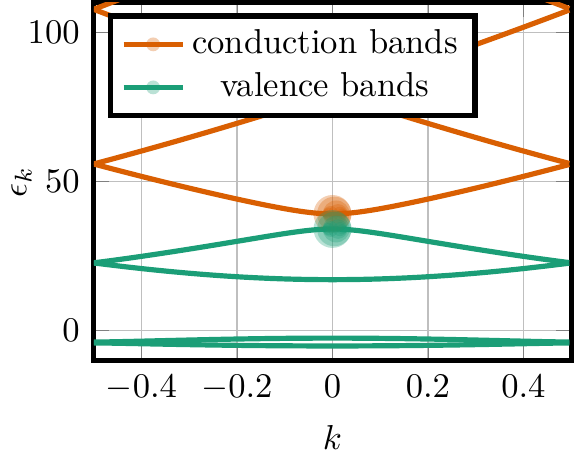}
\caption{On the left: The potentials $V(r,0)$ and $W(r,0)$.
On the right: Band structure with coefficients of the lowest eigen function for $N_k = 128$. The area of the circles on the valence and conduction band at position $\vk$ is proportional to $\sum_{i_c} |X(i_v i_c \vk)|^2$ and $\sum_{i_v} |X(i_v i_c \vk)|^2$, respectively.}
\label{fig:potentialsbands}
\end{figure}

For the one-dimensional problem, we take the single particle orbitals
$\psi_{i\vk}(\vr)$ in \eqref{eq:BSEreal} to be eigenfunctions of a 
single particle Hamiltonian $\mc{H}(\vk)$ in which the effective potential is defined as 
\[
V_\text{eff}(r) = 20 \cos(4 \pi r / L) + 0.2 \sin(2 \pi r / L),
\]
where the unit cell size is $\abs{\Omega}\equiv L = 1.5$. 

The bare Coulomb potential used in \eqref{eq:BSEreal} is chosen to be
\begin{equation}
V(r, r') = \frac{1}{\sqrt{(r - r')^2 + 0.01}},
\end{equation}
and the screened interactions is chosen as
\begin{equation}
W(r, r') = \tfrac{(3 + \sin(2 \pi r / L)) (3 + \cos(4 \pi r' / L))}{16}
e^{-\frac{(r - r')^2}{32 L^2}} V(r, r').
\end{equation}
Compared to the smoothed out Coulomb potential $V$, the chosen screened interaction $W$ decays exponentially and also contains lattice periodic contributions.
The potentials are shown in Figure~\ref{fig:potentialsbands}.
Both potentials are periodically extended $N_k-1$ times outside of the unit cell.
The particular structure of the potentials has an influence on the band structure and spectrum of the BSH, but was observed to not significantly impact the convergence behavior or the runtime scaling of the ISDF method.

The Bloch functions $u_{i\vk}$ are sampled on $N_g = 128$ uniformly distributed 
grid points within the unit cell, and the number of
$\vk$ points $N_k$ ranges from $16$ to $4096$ in our experiments.

For each $\vk$ point, the first four eigenstates are treated as the valence 
states in this model, while the remaining eigenstates are considered as the conduction
states, separated by an energy gap from the former.  
We use all $N_v=4$ valence bands and $N_c = 5$ conduction 
bands to construct the approximate $H_{\textrm{BSE}}$. 
The number of $\vk$ points was chosen to be $N_k = 256$ in the error analysis of the ISDF approximation, and varies from $16$ to $4096$ in the run time analysis and the analysis of the error in the absorption spectrum. The largest resulting Hamiltonian is of size $81920 \times 81920$.

\begin{figure}
  \includegraphics[width=0.48\textwidth]{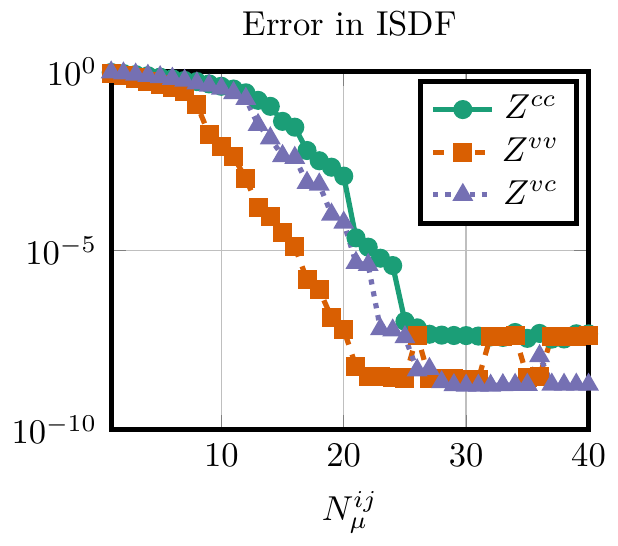}
  \includegraphics[width=0.48\textwidth]{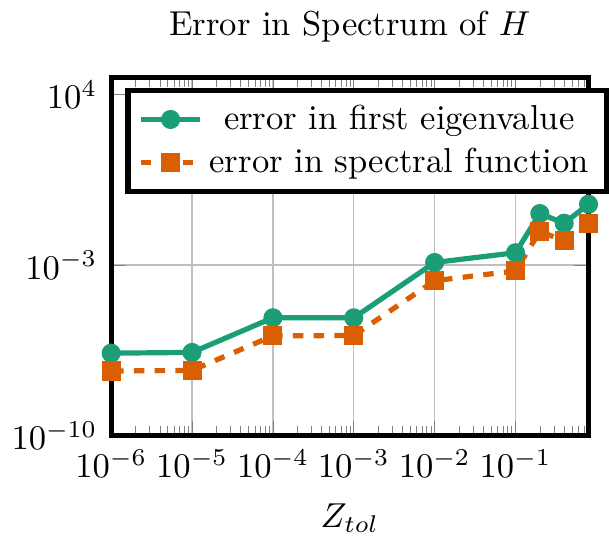}
  \caption{On the left: ISDF approximation error $\lVert Z - \Theta C \rVert_F / \lVert Z \rVert_F$ 
  for different choices of $N_\mu$. On the right: Resulting errors in the spectrum of $H_{\textrm{BSE}}$ for different ISDF error tolerance.}
  \label{fir:1d_isdf_error}
\end{figure}
Figure~\ref{fir:1d_isdf_error} shows how the ISDF approximation error 
varies with respect to the truncation parameter $N_{\mu}^{ij}$ and 
how the accuracy of the approximate spectrum of $H_{\textrm{BSE}}$ changes
with respect to the ISDF approximation error.

In the left subfigure, we plot the relative error $\lVert \Theta^{\alpha\beta} C^{\alpha\beta} -
Z^{\alpha\beta} \rVert_F / \lVert Z^{\alpha\beta} \rVert_F$, $\alpha, \beta \in \{v, c\}$, where $\|\cdot\|_F$ is the
Frobenius norm,  for different choices of truncation levels $N_{\mu}$ (or number of interpolation points).
As expected, when $N_\mu$ is too small, ISDF results in relatively large error. As $N_{\mu}$ becomes slightly larger, the ISDF approximation error 
decays exponentially with respect to $N_{\mu}$ up to $N_{\mu}=20 \sim 30$. 
At this truncation level, the error is on the order of $10^{-8}$,
which is sufficiently small for obtaining an highly accurate approximation
of the spectrum of $H_{\textrm{BSE}}$ as shown in the right subfigure. In this subfigure, we plot the relative error in the first eigenvalue and in the overall optical absorption spectrum against the ISDF error tolerance $Z_{\textrm{tol}}$.  For each $Z_{\textrm{tol}}$, we choose the smallest truncation 
parameters $N_\mu$'s with the resulting error in $Z^{\alpha,\beta}$ being lesser or equal to $Z_{\textrm{tol}}$
for $\alpha, \beta \in \{v,c\}$.

\begin{figure}
\centering
\includegraphics{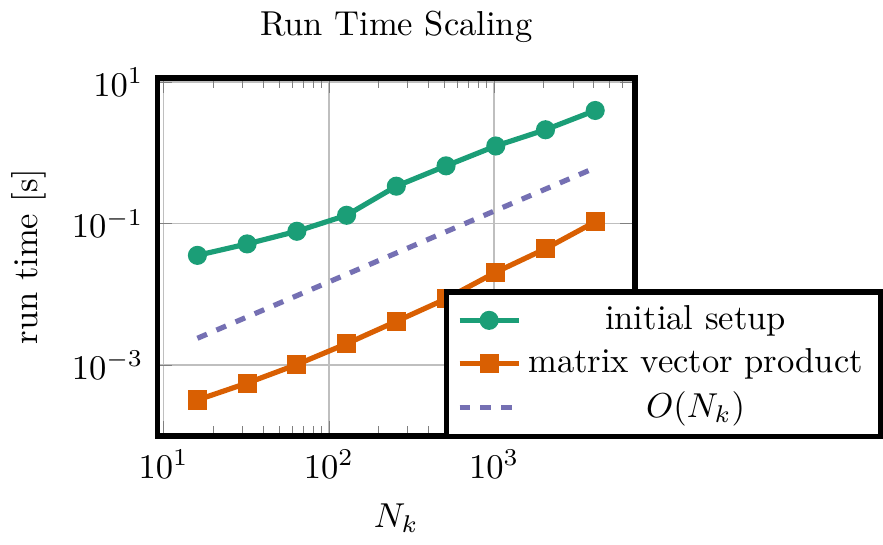}
\caption{Run times for the initial setup and individual matrix-free
matrix-vector products.}
\label{fig:1d_runtimes}
\end{figure}

In Figure~\ref{fig:1d_runtimes}, we plot the timing measurements for 
both the construction of $\wt{V}$ and $\wt{W}$ and the multiplication 
of the approximate $H_{\textrm{BSE}}$ with a vector with respect to $N_k$.
In these calculations, the ISDF truncation parameters $N_\mu$'s are 
chosen so that the relative error in $Z^{\alpha\beta}$ is below 
$Z_{\textrm{tol}}=10^{-5}$. This error tolerance resulted in the choices 
of $N_\mu^{vv} = 17$, $N_\mu^{cc} = 23$, and $N_\mu^{vc} = 21$.

As we can see in Figure~\ref{fig:1d_runtimes}, the scaling of 
the runtime for the construction of  $\wt{V}$ and $\wt{W}$ is nearly
linear with respect to $N_k$, which is in excellent agreement with 
the theoretical computational complexity presented in the preceeding section.
The scaling of the runtime for the multiplication of the approximate
$H_{\textrm{BSE}}$ with a vector also looks linear in $N_k$.
In fact, a more detailed investigation showed that the convolutions in $\vk$
in the application of $W$ dominate the cost of the matrix-vector multiplications,
in good agreement with the theoretical $\Or(N_k \log N_k)$ complexity shown earlier.

For comparison, without the use of ISDF, the construction of 
$H_{\textrm{BSE}}$ is estimated to take about $460,000$ seconds for $N_k = 4096$. With our method it took less than $10$ seconds.

\subsection{Three-dimensional problems}

\begin{figure}
\includegraphics{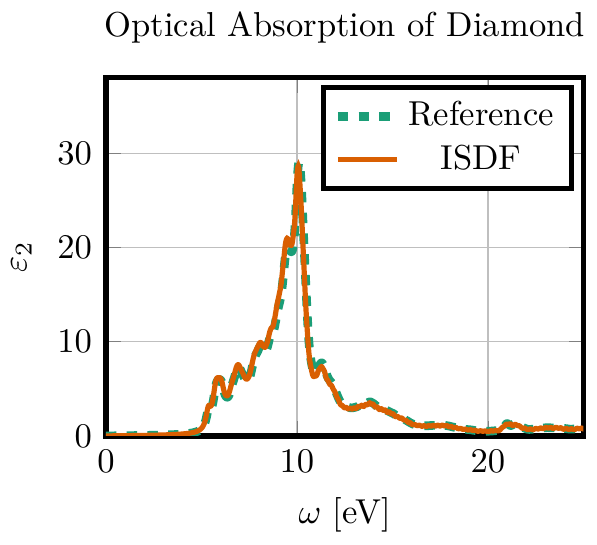}
\includegraphics{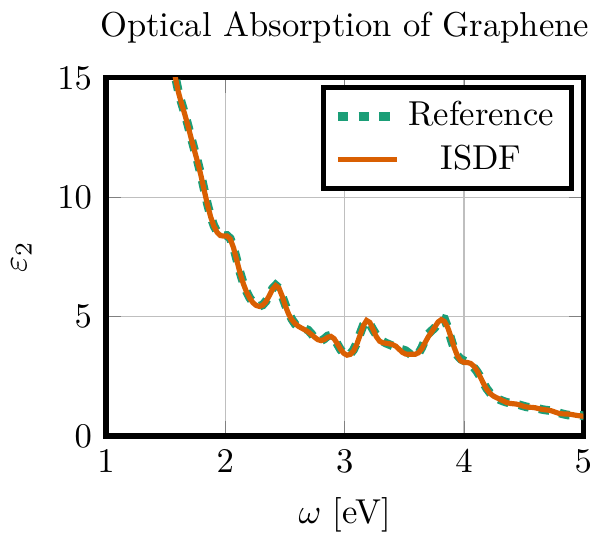}
\caption{Optical absorption spectrum for diamond (left) and graphene (right).}
\label{fig:diamond_graphene_absorption}
\end{figure}

\begin{table}
\begin{center}
\begin{tabular}{lccc}
  \toprule
  Parameters & Diamond & Graphene\\
  \midrule
  $N_v$ & $4$ & $4$\\
  $N_c$ & $10$ & $5$\\
  $N_k$ & $13 \times 13 \times 13$ & $42 \times 42 \times 1$\\
  $N_r$ & $20 \times 20 \times 20$ & $15 \times 15 \times 50$\\
  $N_\mu^{vv}$ & $70$ & $50$\\
  $N_\mu^{cc}$ & $220$ & $180$\\
  $N_\mu^{vc}$ & $100$ & $60$\\
  $N_\text{iter}$ & $150$ & $100$\\
  \bottomrule
\end{tabular}
\end{center}
\caption{Parameters used in the computation of spectra and the benchmarks.}
\label{tab:parameters}
\end{table}
We now compare optical absorption spectra for diamond and graphene computed from the approximate $H_{\textrm{BSE}}$ constructed via ISDF with corresponding reference spectra. The reference spectra are obtained from the exact 
$H_{\textrm{BSE}}$ from the \texttt{exciting} code \cite{exciting,Vorwerk2019}. The comparison is shown in Figure~\ref{fig:diamond_graphene_absorption}. The reference spectrum for diamond is constructed on a $13 \times 13 \times 13$ $\vk$-grid using all 4 valence and 10 conduction states. Fourier components $\hat{W}_{\vq}(\vG,\vG')$ in Eq.~\eqref{eqn:Wfourier} are calculated up to a cut-off $|\vG+\vq|\le 2.5 \; \mathrm{a}_0^{-1}$, where $\mathrm{a}_0$ is the Bohr radius. The screened Coulomb interaction is calculated within the random-phase approximation (RPA) including 100 conduction states. For graphene, the reference spectrum is obtained on a $42 \times 42 \times 1$ $\vk$-grid using all 4 valence and 5 conduction states. Fourier components $\hat{W}_{\vq}(\vG,\vG')$ in Eq.~\eqref{eqn:Wfourier} are calculated up to a cut-off $|\vG+\vq|\le 2.0 \; \mathrm{a}_0^{-1}$ and 80 conduction states are included in the RPA calculations for the screened Coulomb potential.
The numerical parameters of the reference and approximate calculations are shown in Table~\ref{tab:parameters}.
The number of interpolation vectors was chosen such that the relative ISDF error was around $0.1$.

We can clearly see that for both diamond and graphene, the approximate optical absorption spectrum matches well with 
the reference spectrum. In particular, the positions and heights of all major peaks are in good agreement. 
We should note that, in the case of diamond, the absorption spectrum produced by a $13 \times 13 \times 13$ $\vk$-grid is in good 
agreement with measurements \cite{Phillip1964} and previous BSE calculations \cite{Hahn2005}.
In the case of graphene, however, larger $\vk$-grids have been reported for BSE calculations \cite{PhysRevLett.103.186802} to produce
an optical absorption spectrum in good agreement with the experimental result.

\begin{figure}
\centering
\adjustbox{raise=0.8mm}{\includegraphics{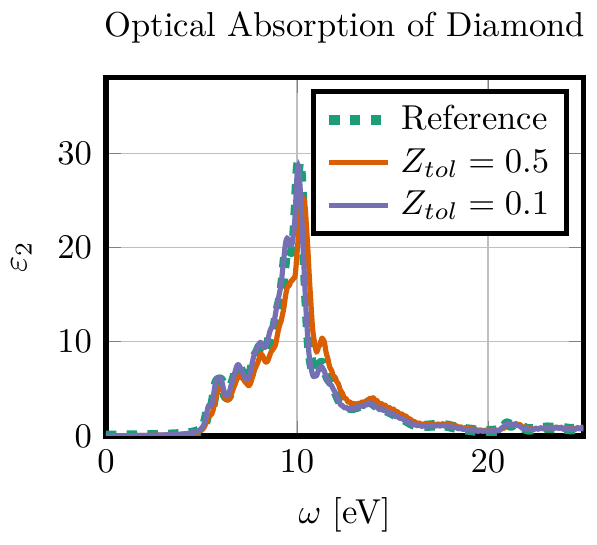}}
\includegraphics{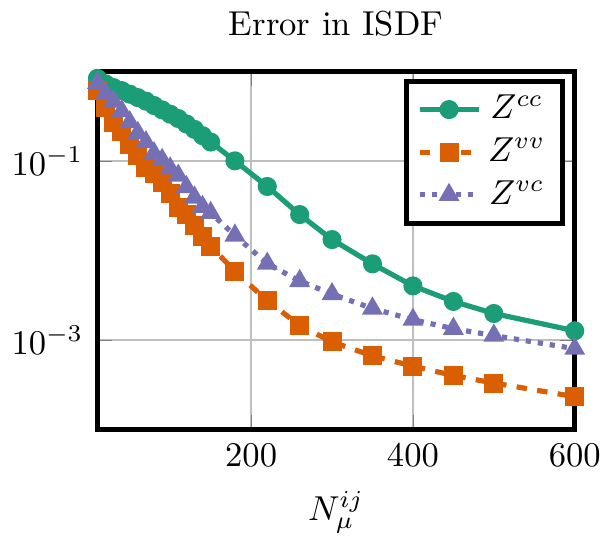}
\caption{On the left: Optical absorption spectrum for diamond with differently accurate ISDF approximations.
On the right: Estimated errors in ISDF approximation with different numbers of interpolation points.}
\label{fig:isdf_error}
\end{figure}
\begin{table}
\begin{center}
\begin{tabular}{lrr}
  \toprule
\multicolumn{3}{c}{Error in}\\
\cmidrule(lr){1-3}
$Z$ & Absorption Function & First Eigenvalue\\
\midrule
$0.5$ & $0.199$ & $0.0038 \,(\SI{20.7}{\milli\eV})$\\
$0.1$ & $0.056$ & $0.0011 \,\,\,\,(\SI{6.2}{\milli\eV})$\\
$0.05$ & $0.040$ & $0.0006 \,\,\,\,(\SI{3.3}{\milli\eV})$\\
\bottomrule
\end{tabular}
\end{center}
\caption{Errors in the spectrum for differently accurate ISDF approximations.}
\label{tab:isdf_error_spectrum}
\end{table}
Figure~\ref{fig:isdf_error} shows that the ISDF approximation error can be systematically reduced as we increase the number interpolating 
vectors $N_\mu$. 
However, Figure~\ref{fig:diamond_graphene_absorption} 
shows that the approximate absorption spectrum is already in 
good agreement with the reference spectrum, when the relative ISDF 
approximation error is at $0.1$. Thus, it seems unnecessary to use a larger number of 
interpolation vectors in these cases. This observation is corroborated by the 
relative difference between the first eigenvalue of the approximate $H_{\textrm{BSE}}$ computed using ARPACK and that of reference
$H_{\textrm{BSE}}$ constructed in \texttt{exciting} shown in Table~\ref{tab:isdf_error_spectrum}.
With a relative ISDF approximation error of $Z_{tol}=0.1$, the error in the first BSE eigenvalue is below $10 \; \mathrm{meV}$ in both examples 
shown here.

\begin{figure}
\centering
\includegraphics{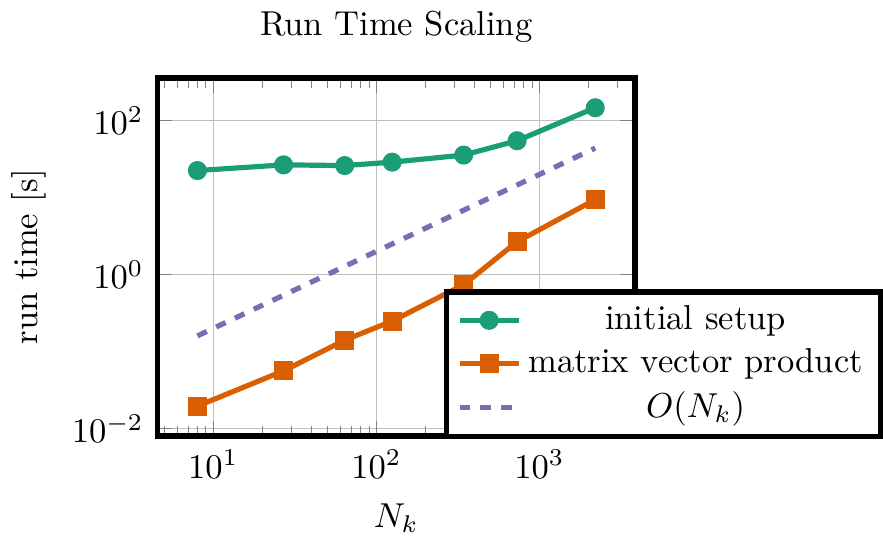}
\caption{Run times for the initial setup and individual matrix-free
matrix-vector products.}
\label{fig:diamond_runtimes}
\end{figure}
To illustrate the run time scaling of the method in the 3D examples, we measure the time it takes to construct the approximate $H_{\textrm{BSE}}$ via ISDF as 
well as the time it takes to multiply the resulting $H_{\textrm{BSE}}$ with vectors for the diamond example.
We use $\vk$-grids of sizes $N_k = n_k \times n_k \times n_k$ for $n_k \in \{2, 3, 4, 5, 7, 9, 13\}$. The resulting timing measurements are 
plotted in Figure~\ref{fig:diamond_runtimes}.
It can be seen that the runtime for constructing the approximate $H_{\textrm{BSE}}$ scales linearly with the number of $\vk$-points. 
The multiplication 
of $H_{\textrm{BSE}}$ with vectors scales as $\mc{O}(N_k \log(N_k))$ for 
sufficiently large $N_k$.
As in the model problem, the convolutions in $\vk$ in the application of $W$ dominate the cost of the matrix-vector multiplications.
For comparison, computing the ISDF decomposition of the Hamiltonian for the case $N_k =
13^3$ took $147$ seconds, whereas the full assembly of the Hamiltonian took about 6 hours in \texttt{exciting} on 13 compute nodes with
13 cores each. The optical absorption function was obtained by running
about $150$ Lanczos steps, which amounts to about $24$ minutes for each fixed direction (x, y, and z), compared to almost 4 hours required in the \texttt{exciting} code for the full diagonalization on 13 compute nodes.

\section{Conclusion}\label{sec:conclude}
In this paper, we examined the possibility of using the ISDF technique
to reduce the computational complexity of BSH construction and the 
subsequent iterative approximation of 
the optical absorption spectrum and excitation energies of electron-hole
(exciton) pairs for solids. For periodic systems, a fine $\vk$-point sampling
in the Brillouin zone is often required to produce accurate 
results, whereas the number of bands per $\vk$-point 
required to construct the bare exchange and screened direct kernels of the BSH  
is relatively small.  We showed that the complexity 
of the ISDF procedure scales linearly with respect to the number
of $\vk$ points ($N_k$) when the ranks of the approximate 
bare exchange and screened direct kernels produced by the ISDF 
procedure are chosen to be independent of $N_k$.
By keeping the bare exchange and screened direct kernels in the 
low-rank decomposed form produced by the ISDF procedure, an iterative
method used to obtain the optical absorption spectrum and selected 
excitation energies (eigenvalues of the BSH) can be implemented
with cost scaling as $\Or(N_k \log N_k)$.
Our numerical experiments, which were performed on a 1D model
as well as two different types of actual materials (diamond and graphene), 
confirm our complexity analysis. They demonstrate that the ISDF technique
can indeed significantly reduce the cost of BSE calculation for 
solids while maintaining the same accuracy provided by a standard 
BSE calculation implemented in the software \texttt{exciting}. 
Our current implementation of the ISDF technique is done using the \texttt{Julia} programming language for a single node. A distributed parallel 
implementation is needed to accommodate a much finer $\vk$-point 
sampling which is required in case of the graphene example to produce 
a computed absorption spectrum that matches with experimental results.

\section*{Acknowledgments}

This work was partially supported by the Department of Energy under grant DE-SC0017867 (L.L.),  by the Center for Computational Study of Excited-State Phenomena in Energy Materials (C2SEPEM) at the Lawrence Berkeley National Laboratory, which is funded by the U.\,S. Department of Energy, Office of Science, Basic Energy Sciences, Materials Sciences and Engineering Division, under Contract No. DE-AC02-05CH11231 (C.Y.), by the Scientific Discovery through Advanced Computing (SciDAC) program, and by the CAMERA program (L.L. and C.Y.). Within a framework cooperations between the University of California at Berkeley and Freie Universit\"at Berlin, the latter sponsored an extended visit of F.H.\ and R.K.\ in Berkeley. We thank Wei Hu, Meiyue Shao and Kyle Thicke for helpful discussions. C.D. and R.K. thank IPAM, UCLA, for its support during the 2013 fall program on ``Materials for a sustainable energy future'' and for creating the inspiring scientific atmosphere that initiated their collaboration.

\bibliographystyle{siam}
\bibliography{bseisdf}

\end{document}